\documentclass[pre,superscriptaddress,showpacs,eqsecnum,twocolumn]{revtex4}
\usepackage{newlfont}
\usepackage{amssymb}
\usepackage{amsfonts}
\usepackage{amsmath}
\usepackage{graphicx}
\usepackage{bm}
\begin{document}
\title{Global entangling properties of the coupled kicked tops}
\author{Rafa{\l} Demkowicz-Dobrza\'nski}
\affiliation{Center for Theoretical Physics, Polish Academy of Sciences\\
Aleja Lotnik{\'o}w 32/44, 02-668 Warszawa, Poland}
\author{Marek Ku{\'s}}
\affiliation{Center for Theoretical Physics, Polish Academy of Sciences\\
Aleja Lotnik{\'o}w 32/44, 02-668 Warszawa, Poland}
\affiliation{Faculty of Mathematics and Sciences, Cardinal Stefan Wyszy\'nski University, Warszawa, Poland}
\date{\today}

\begin{abstract}
We study global entangling properties of the system of coupled kicked tops
testing various hypotheses and predictions concerning entanglement in quantum
chaotic systems. In order to analyze the averaged initial entanglement
production rate and the averaged asymptotic entanglement various ensembles of
initial product states are evolved. Two different ensembles with natural
probability distribution are considered: product states of independent
spin-coherent states and product states of random states. It appears that
the choice of either of these ensembles results in significantly different
averaged entanglement behavior. We investigate also a relation between the
averaged asymptotic entanglement and the mean entanglement of eigenvectors of
the evolution operator. Lower bound on the averaged asymptotic entanglement is
derived, expressed in terms of the eigenvector entanglement.

\end{abstract}

\maketitle

\section{Introduction}
Looking for quantum signatures of the classical transition from regular to
chaotic dynamics is the field of quantum chaos \cite{haake,reichl}. Recently it
was suggested \cite{furuya1997,miller1998} that entanglement production in a
quantum system can be a good indicator of the regular to chaotic transition in
its classical counterpart.

In both of the invoked studies \cite{furuya1997,miller1998} it was observed
that the presence of chaos enhances the rate at which an initial product state
is getting entangled. In the paper \cite{miller1998} Miller and Sarkar
considered coupled quantum kicked tops. Single kicked top is a thoroughly
studied model in the quantum chaos literature. Depending on the strength of
kicks its classical dynamics is either regular or chaotic. Miller and Sarkar
studied a system consisting of two identical kicked tops with an additional
weak interaction between them. The strength of kicks was chosen in such a way
that the classical phase space of a single top was mixed (consisted both of
regular tori and chaotic regions). Two tops were initially in a product state
of two spin-coherent states. The reason for this choice is that spin-coherent
states have a good classical limit, what gives a chance to relate the classical
phase space picture to the quantum description. The spin-coherent state of the
first top was chosen to lie in a chaotic region while the spin-coherent state
of the second one was varied from a regular to a chaotic region. The system was
then evolved and it was observed that the rate of entanglement increase
(measured as the von Neumann entropy of the reduced density matrix) was higher
when the second top was placed in a chaotic region. More quantitatively, it was
shown that for different quantum initial states the rate of entanglement
increase was proportional to the sum of two positive Lyapunov exponents
calculated for the corresponding classical distribution of initial points.
These results supported the claim about close relation between chaotic behavior
of classical systems and entangling properties of their quantum versions.

Further investigations revealed, however, that there is no such direct relation
between chaos and entanglement
\cite{angelo1999,tanaka2002,fujisaki2003,prosen2002,prosen2003,znidaric2003,wang2003}.
In particular it was observed that it is rather a specific time correlation
function than the Lyapunov exponent itself that determines the entanglement
production rate. Strong chaos is able to destroy the time correlations even
on a very short time scale and thus it diminishes initial entanglement
production rate.

Apart from studying the entanglement production rate, which is a quantity
calculated from a short-time behavior of the evolved state, one can also study
asymptotic properties of entanglement, i.e.\ those appearing in the long-time
limit. The question whether the asymptotic value of entanglement is related to
chaos was posed in \cite{bandyopadhyay2003}. If the coupling strength between
kicked tops is large enough to observe the saturation of entanglement within
given time, the asymptotic value of entanglement is higher the greater is the
chaos parameter (the kick strength). This observation was made starting with an
initial state in form of a product of two spin-coherent states placed in a
region which for low kick strengths was regular, contained a part of a
separatrix for stronger kicks, and eventually became chaotic for very strong
kicks.

In the present paper we also concentrate on the case of coupled kicked tops. In
most of the previous studies the attention was focused on analyzing how an
initial product state of certain spin-coherent states evolves in time. Here we
would like to address a more general problem. Namely, how does the strength of
chaos influence the global entangling properties of the evolution of the kicked
tops. By global we mean properties not depending on a specific choice of
initial product state. In other words this is the question of whether the
entangling capabilities of the transformation depend on the strength of chaos.
Global entangling properties were analyzed in the kicked Ising spin chain
model \cite{prosen2002}, where evolution of random states was investigated, and
decrease of entanglement production rate with the increase of chaos was
observed.

Another approach often used \cite{lakshminarayan2001} in determining the
entangling power of an operation consists of analyzing its eigenvectors. In the
case of periodically driven systems these are the eigenvectors of an unitary
operator $U$ corresponding to the one-period evolution. The degree of
eigenvector entanglement is then regarded as an information about entangling
properties of the evolution. We show that the information about entanglement of
eigenvectors does not, however, give the full picture of entanglement
properties of the evolution, but only a rough estimate of the asymptotic
behavior of entanglement of evolved states (see Sec.\ref{sec:eigenasymp}).

In order to discuss entangling properties of the evolution of the kicked tops,
we evolve not a single product state, but a whole ensemble of random product
states, chosen uniformly with respect to action of $\textrm{SU}(d_1)\otimes
\textrm{SU}(d_2)$ group, where $d_1$, $d_2$ are the Hilbert space dimensions of
the two tops. We calculate the averaged asymptotic entanglement approached by
evolved states, and the initial rate of entanglement production. All pure
states are treated on equal footing here. Spin coherent states are as good as
any other pure state. This way of averaging gives us information about
\emph{the entangling power of evolution}, as defined in \cite{Zanardi2000}.

Additionally we shall calculate the entanglement production when initial states
are products of two spin-coherent states, where each spin-coherent state,
parameterized by two spherical angles $\theta$, $\phi$, is taken independently
from the ensemble with uniform probability density on the unit sphere. The two
ways of averaging give qualitatively different results. We shall analyze the
results with the help of the perturbative approach
\cite{fujisaki2003,znidaric2003} and the analysis of the entanglement of
eigenvectors.

The observation that is worth mentioning here is that even in a very regular
regime \emph{the entangling power} is extremely high -- higher than in chaotic
cases -- both when discussing the asymptotic entanglement behavior and the
initial entanglement production rate.

Our results are another step to reveal the relation between chaotic vs. regular
motion and entanglement production.

\section{Entanglement}
Entanglement is a purely quantum phenomenon, dividing states of a composite
quantum systems into two classes: those which can be written as products of
some states of the (two in this case) subsystems:
\begin{equation}\label{product}
|\Psi\rangle=|\psi_1\rangle\otimes|\psi_2\rangle,
\end{equation}
called product states, and all other which cannot be written in the form
(\ref{product}) but instead involve a genuine, nontrivial linear combination
\begin{equation}\label{entanglement}
|\Psi\rangle=\sum_{ij}c_{ij}|\psi_{1i}\rangle\otimes|\psi_{2j}\rangle.
\end{equation}
The definition above applies only when one deals with pure states. The notion
of entanglement for mixed states is more subtle \cite{werner1989}, and it is
very difficult in general to determine whether a given mixed state is entangled
or not. In this paper the state of the composite quantum system we consider is
always pure.

For a given pure state $|\Psi\rangle$ it is easy to check whether it is
entangled or not. Observe that if the state is a product one, averaging over
one of the subsystems of the corresponding density matrix
$\rho=|\Psi\rangle\langle\Psi|$ gives a pure state of the remaining subsystem:
\begin{equation}\label{tr2}
\rho_1:=\mathrm{Tr}_2(\rho)=|\psi_1\rangle\langle\psi_1|,
\end{equation}
whereas the same procedure applied to an entangled state produces necessarily a
genuine mixed state. This observation can be further quantified by calculating
the linear entropy for the reduced density matrix
\begin{equation}\label{linentr}
  S_L:=1-\mathrm{Tr}(\rho_1^2),
\end{equation}
which vanishes for product states and reaches the maximal value:
\begin{equation}
  \label{eq:Smax}
S_{L}^{\textrm{max}}=1-1/d,
\end{equation}
where $d=\textrm{min}(d_1,d_2)$ is the Hilbert space dimension of the smaller
subsystem, for the ''maximally entangled state'' (by definition this is the state
which reduces to the ''maximally mixed'' state of a subsystem -- the latter is
characterized by a diagonal density matrix with equal entries). In this sense
$S$ is a measure of entanglement for pure states of a composite system.

Another often used measure of entanglement for pure states is the von Neumann
entropy of the reduced density matrix:
\begin{equation}
\label{eq:vonNeumann}
S_{\textrm{vN}} = -\textrm{Tr}\left(\rho_1 \log_2\rho_1\right).
\end{equation}
For product states $S_{\textrm{vN}}=0$, as the reduced density matrix $\rho_1$
is also pure, while for the maximally entangled states the von Neumann entropy
of the reduced density matrix takes the highest value
$S_{\textrm{vN}}^\textrm{max}=\log_2d$. Contrary to the linear entropy, the
above measure of entanglement has a nice operational meaning in terms of the
number of maximally entangled states that can be distilled from a given number
of non-maximally entangled states \cite{bennett1989}.

Actually all quantities which do not increase under local operations (ie.\
operations acting separately in each subsystem) and classical communication
quantify in some way the amount of entanglement present in a state. These in
general are called \emph{entanglement monotones} \cite{vidal2000}. Linear and
von Neumann entropy discussed above are examples of such entanglement monotones
for pure states.

In the following we shall use the linear entropy (\ref{linentr}) as the measure
of entanglement. We choose this measure, instead of the von Neumann entropy
(\ref{eq:vonNeumann}), as linear entropy is easier to calculate and there is a
perturbative formula for initial growth of linear entropy
\cite{fujisaki2003,znidaric2003} in weakly coupled systems which we shall use.
Furthermore, in the investigations concerning relation between chaos and
entanglement, where both von Neumann and linear entropy where calculated
\cite{bandyopadhyay2003,fujisaki2003,bandyopadhyay2002} no qualitative
difference in the behavior of the two was found, thus the choice of either of
them is not crucial.

It is argued that the presence of entanglement is important in many novel
applications of quantum information processing \cite{nielsen}, which explains the
prominence enjoyed by this phenomenon in many recent investigations. In our
study we shall concentrate only on the interplay between production of
entanglement in a composite quantum system and its chaotic properties.

\section{Coupled kicked tops}
The kicked top is a paradigmatic model for studying quantum chaos in
finite-dimensional Hilbert spaces \cite{hks87a}. It is a particle with the
total spin $j$ and the dynamics generated by the Hamiltonian
\begin{equation}\label{kt1}
  H=pJ_y+\frac{k}{2j}J_z^2\sum_{n=-\infty}^\infty\delta(t-n).
\end{equation}
Here $J_y$ and $J_z$ are the components of the angular momentum operator
fulfilling the standard commutation relations $[J_y,J_z]=iJ_x$, etc. The
time dependence takes the form of an infinite train of delta-shaped pulses
(''kicks'') perturbing the free rotation periodically in time. The quantities $p$
and $k$ are adjustable parameters of the model. The latter, called the
kick-strength, is scaled by the total spin $j$ -- observe that the total angular
momentum $J^2=J_x^2+J_y^2+J_z^2$ is conserved, $[J^2,H]=0$, hence we can
investigate the dynamics of the model for each value of $j$ independently,
restricting the discussion to the appropriate $\left[(2j+1)\times (2j+1)\right]$-dimensional
space. To exhibit various interesting dynamical aspects of the model it is
enough to change one of the parameters. In the following we put $p=\pi/2$ and
vary $k$.

The unitary time evolution operator transporting the wave function of the
kicked top in time over one period of the perturbation,
\begin{equation}\label{flo1}
U=\exp\left(-i\frac{k}{2j}J_z^2\right)\exp\left(-i\frac{\pi}{2}J_y\right),
\end{equation}
generates the Heisenberg equations of motion for the angular momentum
operators $J_x, J_y$, and $J_z$
\begin{equation}
\label{qem1}
\begin{array}{rcl}
  J_x^\prime&=&U^\dagger J_xU=\frac{1}{2}\left(J_z+iJ_y\right)e^{-i(k/j)\left(J_x-1/2\right)}
  +\mathrm{h.c.}, \\
  J_y^\prime&=&U^\dagger J_yU=\frac{1}{2i}\left(-J_z+iJ_y\right)e^{-i(k/j)\left(J_x-1/2\right)}
  +\mathrm{h.c.}, \\
  J_z^\prime&=&U^\dagger J_xU=-J_x,
\end{array}
\end{equation}
giving the operators $J_x^\prime,J_y^\prime$, and $J_z^\prime$ at time $t=n+1$ in
terms of their predecessors $J_x,J_y$, and $J_z$ at time $t=n$.

As in all studies of quantum chaotic phenomena we are ultimately interested in
comparing quantum and classical dynamics of the model. In the present the
Planck constant has been set to unity hence the classical limit corresponds to
$j\to\infty$ limit (''large quantum numbers''). More formally one defines the
quantities $X,Y$ and $Z$ as averages of $J_x/j,J_y/j$, and $J_z/j$ calculated
in the initial state of the system. In comparing classical and quantum behavior
it is reasonable to take as an initial state some minimum uncertainty state, in
belief that (at least in the large $j$ limit) the evolution of averages will be
well represented by the single classical trajectory starting from the point in
the phase space around which the initial quantum state of minimal uncertainty
is concentrated. Obviously, there is no particular reason to distinguish such
states when investigating purely quantum properties like entanglement
production. Appropriate minimum uncertainty states for spin $j$ particles (so
called angular momentum coherent states) can be generated from the
$\left|j,j\right\rangle$ state [i.e. the common eigenstate of $J_z$ and $J^2$
with the eigenvalues $m=j$ and $j(j+1)$ respectively] by unitary rotations
\begin{equation}\label{cs}
\left|\theta,\phi\right\rangle=\left(1+|\gamma|^2\right)^{-j}
e^{\gamma(J_x-iJ_y)}\left|j,j\right\rangle, \quad
\gamma=e^{i\phi}\tan\frac{\theta}{2},
\end{equation}
and the above described procedure leads to the following classical mapping:
\begin{equation}\label{cem1}
\begin{array}{rcl}
X^\prime&=&Z\cos(kX)+Y\sin(kX), \\
Y^\prime&=&-Z\sin(kX)+Y\cos(kX), \\
Z^\prime&=&-X.
\end{array}
\end{equation}
A detailed analysis of the classical dynamics (\ref{cem1}) is given in
\cite{hks87a}, let us only summarize that the system is integrable for $k=0$
and becomes visibly chaotic when $k > 2$. For $k$ around $3$ the phase space
exhibits well developed mixed structure with few regular islands embedded in
the chaotic see. When $k\approx 6$ islands of stability, although present, are
very small and the chaos can be treated as fully developed for all practical
purposes. From the point of view of quantum chaos, the islands are negligible
if their phase-space area is smaller then the effective value of the Planck
constant ($1/j$ in our case), which will be the case in our calculations.

In order to achieve our ultimate goal, ie.\ the investigation of parallels
between chaos and entanglement we follow the idea of Miller and Sarkar
\cite{miller1998} and consider two coupled kicked tops with the Hamiltonian
\begin{equation}\label{kt2}
  H=H_1+H_2+H_I,
\end{equation}
where $H_1$ and $H_2$ are the Hamiltonians of independent kicked tops
(\ref{kt1}) expressed in terms of the operators $J_{x_1}, J_{y_1}, J_{z_1}$
and $J_{x_2}, J_{y_2}, J_{z_2}$ pertaining to each individual top, whereas
$H_I$ is a nonlinear coupling term
\begin{equation}\label{hi}
  H_I=\frac{\epsilon}{j}J_{z_1}J_{z_2}\sum_{n=-\infty}^\infty\delta(t-n).
\end{equation}
The procedure of obtaining the classical evolution equations is exactly the
same as the one for a single top described above, and yields
\cite{bandyopadhyay2003}:
\begin{equation}
\begin{array}{rcl}
X_1^\prime &=& Z_1 \cos \Delta_{12}+Y_1 \sin \Delta_{12} \\
Y_1^\prime &=& -Z_1 \sin \Delta_{12}+Y_1 \cos \Delta_{12} \\
Z_1^\prime &=& -X_1\\
X_2^\prime &=& Z_2 \cos \Delta_{21}+Y_2 \sin \Delta_{21} \\
Y_2^\prime &=& -Z_2 \sin \Delta_{21}+Y_2 \cos \Delta_{21} \\
Z_2^\prime &=& -X_2,\\
\end{array}
\end{equation}
where $\Delta_{12}= kX_1+\epsilon X_2$, $\Delta_{21}= kX_2+\epsilon X_1$.
In the most of the following the coupling strengths $\epsilon$ will be small
in comparison with $k$, it means that the degree of chaos in the system is
determined solely by properties of dynamics of individual tops.

\section{Entangling power}

The main idea behind quantifying the entangling power of quantum evolution is
to measure the ability to produce an entangled state out of an initial product
state in the course of the quantum evolution. Although for particular reasons
or applications we can choose a concrete initial state and follow evolution of
its entanglement properties when time passes, such a history would definitely
bear a lot of imprints of the initial state we chose to start with. In our
study we are more interested in entanglement capabilities of the system itself,
so it is more reasonable to take the average over some set of initial states --
the idea advanced by Zanardi \cite{Zanardi2000}
\begin{equation}\label{epower}
  e_p(U)=\left\langle\left\langle
S[U\left(\left|\psi_1\right\rangle\otimes\left|\psi_2\right\rangle\right)]
\right\rangle\right\rangle_{\psi_1,\psi_2}.
\end{equation}
In the above formula $S$ is some appropriate measure of entanglement (in our
case it will be the linear entropy), and $\langle\langle \cdot
\rangle\rangle_{\psi_1,\psi_2}$ denotes averaging over a set of initial product
states $\left|\psi_1\right\rangle\otimes\left|\psi_2\right\rangle$.

The averaging procedure, however, needs more detailed reflection. As mentioned
in the previous section, when investigating the quantum-classical
correspondence, it is reasonable to take as an initial state a spin-coherent
state and, consequently, a product of two such states for a composite system.
The averaging amounts to integrating over the whole set of spin-coherent states
parametrized by the spherical angles $\theta_i$ and $\phi_i$, $i=1,2$ in
(\ref{cs}). In order not to distinguish any particular initial state, we
average over products of two spin-coherent states each parameterized by
$\theta_i$ and $\phi_i$ taken independently from the ensemble with the uniform
probability density on the unit sphere. This kind of averaging will be denoted
as $\textrm{SU}(2)\times \textrm{SU}(2)$ averaging, as our ensemble is
invariant under the action of rotation in either of subsystems.

As already written, there is no particular reason for such a choice of the set
of initial states when global entangling production properties of our system
are investigated. Instead one can average over the whole set of initial product
states. In order to treat all pure product states on equal footing one should
choose the ensemble of product states with probability distribution invariant
under the action of $\textrm{SU}(d_1) \times \textrm{SU}(d_2)$, where $d_1$,
$d_2$ are the Hilbert space dimensions for the subsystems. In this way we
obtain a natural ensemble of random product states.

It is not a surprise that $\textrm{SU}(2)\times \textrm{SU}(2)$ and
$\textrm{SU}(d_1)\times \textrm{SU}(d_2)$ averages can lead to different
quantitative estimates of the entangling power; what is more important they
differ also qualitatively.

One can also think about other characterizations of entangling capabilities of
quantum evolution operators. For example, entanglement properties of
eigenvectors of $U$ can give some information about possible entanglement
production. The matter, however, is rather subtle, as it will be clear from the
subsequent discussion.

\section{Numerical results for coupled kicked tops}

In this section we present the main results of the numerical calculations of
the evolution of the coupled kicked tops. The spins of our two tops are
respectively $j_1=19.5$ and $j_2=20$. This choice is an effect of a compromise.
The spins should be high enough to allow classical correspondence and low
enough to be numerically tractable. Other authors, who considered higher spins
($j = 40$, $j=80$), were able to perform their calculations as they were
evolving only a few different states. Our calculation are performed on
ensembles consisting of several hundreds of states. Consequently we sacrifice
the spin magnitude for the sake of being able to perform averaging over many
states. The chosen spins $j_1$, $j_2$ are not equal. This does not change
evolution significantly (as compared with $j_1=j_2=20$), but removes degeneracy
among eigenstates of one-period evolution operator $U$. Lack of degeneracy is
essential for the eigenvectors entanglement analysis, which will be explained
later.

The coupling strength is chosen to be $\epsilon=0.01$, while the strength of
kicks $k$ (equal for both tops) will be varied from $k=0$ to $k=6$. Chaos
enters the classical dynamics of a single top at $k\approx 2$. The coupling
constant is small enough to assure that all chaotic behavior is due to the
kicks of the tops and not their interaction.
\subsection{Entanglement evolution}
\begin{figure}[t]
\begin{center}
\includegraphics[width=0.5\textwidth]{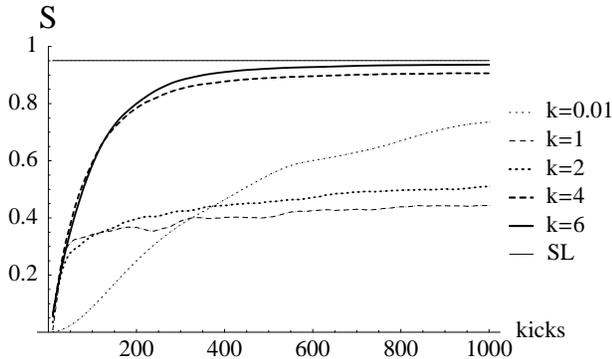}
\caption{The time evolution of the linear entropy averaged over initial
spin-coherent product states ($S_{\textrm{SU}(2)}$), calculated for different
chaoticity parameters $k$ of subsystems (different kick strengths). Spin
magnitudes of two interacting tops are taken respectively $j_1=19.5$, $j_2=20$
and the coupling strength $\epsilon=0.01$. The evolution time comprises $1000$
kicks. The statistical limit (SL) for entanglement is drawn with the solid line
and corresponds to the average entanglement of a randomly chosen state of the
composite system. Increasing the chaoticity parameter causes in general an
increase in the asymptotic value of entanglement (with the exception of the
case $k=0.01$). Initial entanglement growth is extremely slow for very regular
dynamics ($k=0.01$)}
 \label{evolution:spin}
\end{center}
\end{figure}

\begin{figure}[t]
\begin{center}
\includegraphics[width=0.5\textwidth]{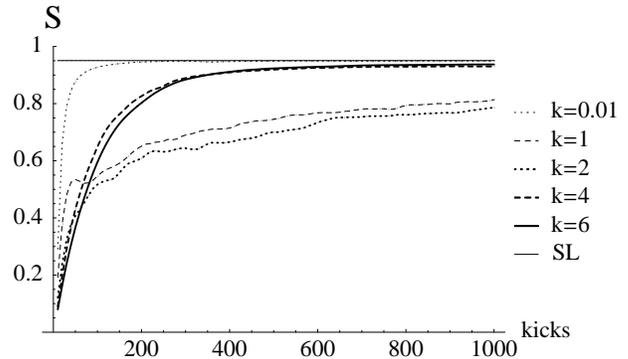}
\caption{The time evolution of the linear entropy averaged over the ensemble of
initial random product states ($S_{\textrm{SU}(d)}$), with the probability
distribution invariant under the action of the $\textrm{SU}(d_1) \times
\textrm{SU}(d_2)$ group, where $d_1=2 j_1+1$, $d_2=2 j_2+1$ are the dimensions
of the subsystems. Different curves correspond to different chaoticity
parameters $k$ of the subsystems (different kick strengths). Spin magnitudes of
two interacting tops are taken respectively $j_1=19.5$, $j_2=20$ and the
coupling strength $\epsilon=0.01$. The evolution time comprises $1000$ kicks.
The statistical limit (SL) for entanglement is drawn with the solid line and
corresponds to the average entanglement of a randomly chosen state of the
composite system. The initial entanglement growth is higher for regular
dynamics (low $k$) than for the chaotic one (high $k$). While asymptotic values
of entanglement are high for chaotic cases, the asymptotic value for the most
regular dynamics $k=0.01$ outperforms all other cases.} \label{evolution:full}
\end{center}
\end{figure}

In Fig.~\ref{evolution:spin} and Fig.~\ref{evolution:full} we show calculations
for $1000$ iterations of tops evolution. In Fig.~\ref{evolution:spin} the
evolution of linear entropy averaged over $100$ random initial spin-coherent
product states is presented.

During the first $1000$ kicks most of the curves saturate to some asymptotic
value. The only exception,  $k=0.01$, requires a little more time to saturate.
The saturation of entanglement in the evolution of the kicked tops is present
also in the case of single spin-coherent product state evolution -- the
averaging over spin-coherent states is not necessary, yet the averaged curves
have smoother behavior. Different curves saturate to different asymptotic
values. This observation was made in \cite{bandyopadhyay2003}, where it was
pointed out that higher is the chaos parameter $k$, the higher is also the
asymptotic value of entanglement reached by the spin-coherent product state
$|\theta,\phi \rangle \otimes |\theta, \phi \rangle$ ($\theta=0.89,
\phi=0.63$). Our averaged results confirm this observation only partially.
While the asymptotic value indeed increases with $k$ for $k \geq 1$, it is also
quite high in the nonchaotic regime - $k=0.01$. For $k=0.01$ the asymptotic
value is higher than that for $k=1$, $k=2$. Opposite to the nonmonotonic
behavior of the asymptotic values, the initial entanglement production rates
seem to increase monotonically with $k$. We shall discuss these observation
more thoroughly in the following. The statistical limit (SL) calculated as an
average entanglement of a random pure state of the full system
\cite{zyczkowski2001} is drawn with a thin solid line.

In Fig.~\ref{evolution:full} we present the result of averaging over $100$
random product states with the probability distribution invariant under the
action of the $\textrm{SU}(d_1) \times \textrm{SU}(d_2)$ group, where $d_1=2
j_1+1$,$d_2=2 j_2+1$ are the dimensions of subsystems (the state of one top is
chosen independently from the state of the other). We generated random product
states using random unitary matrices, distributed uniformly according to the
$\textrm{SU}(d_1) \times \textrm{SU}(d_2)$ Haar measure, applied to a fixed
product state \cite{pozniak2000}. Different asymptotic values of entanglement
for different $k$ is again visible. A monotonic increase of the asymptotic
values can be observed for $k=2,4,6$. However, the differences in the
asymptotic values for $k=4,6$ are tiny. The $k=0.01$  case is especially
interesting. During $1000$ kicks the entanglement saturates to an extremely
high value -- higher than for all other values of $k$. Again, a nonmonotonic
$k$-dependence of the asymptotic entanglement is observed.

Studying the initial production rate of entanglement one can observe
anticorrelation with the parameter $k$. The fastest initial growth of
entanglement corresponds to $k=0.01$, and the slowest to highly chaotic cases
$k=4,6$ -- chaos suppresses the initial entanglement production rate.

For shortening the notation we shall denote the $\textrm{SU}(2) \times
\textrm{SU}(2)$ averaging by the $\textrm{SU}(2)$ averaging and the
$\textrm{SU}(d_1)\times \textrm{SU}(d_2)$ by the $\textrm{SU}(d)$ averaging.
The $\textrm{SU}(d)$ averaged behavior of initial entanglement growth is
strikingly different as compared with the $\textrm{SU}(2)$ averaging. In the
latter case the initial entanglement production rate was almost zero for low
values of $k$, while in the former it was extremely high. Consequently, one
should always distinguish between the entangling power of an evolution and its
particular entangling properties in acting on a certain group of states as
these two may behave very differently.

Summing up the qualitative discussion we conclude that the entangling power of
the evolution [corresponding to $\textrm{SU}(d)$ averaging] both in terms of
the asymptotic value and the initial growth is the highest for very low $k$ --
i.e.\ for very regular dynamics.

\subsection{Asymptotic behavior}

\begin{figure}[t]
\begin{center}
\includegraphics[width=0.5\textwidth]{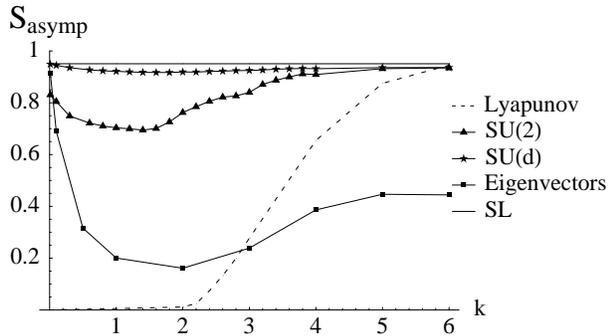}
\caption{This figure presents the dependence of asymptotic values of entanglement
on the chaoticity parameter $k$. Asymptotic values of entanglement
corresponding to averaging over initial random product states [the
$\textrm{SU}(d)$ averaging] are denoted by stars and asymptotic values
corresponding to the averaging over initial spin-coherent product states [the
$\textrm{SU}(2)$ averaging] are denoted by triangles. The dashed line
represents the Lyapunov exponent of the classical dynamics of a single top. The
mean entanglement of eigenvectors of the evolution is denoted by rectangles.
Statistical limit (SL) for entanglement is drawn with the solid line and
corresponds to average entanglement of a randomly chosen state of the composite
system. Chaos indeed increases asymptotic value of entanglement in the case of
spin-coherent states and, to some extent, also in the case of random product
states; nevertheless very regular regime $k\approx 0$ also manifests high
asymptotic entanglement which can be related to high entanglement of
eigenvectors. } \label{comparison:asymptotic}
\end{center}
\end{figure}

We give here more detailed results on asymptotic entanglement. In
Fig.~\ref{comparison:asymptotic} the asymptotic values of entanglement are
presented for different values of $k$. For more credible results the asymptotic
values were obtained as averages of the linear entropy over the evolution of
the tops between 50,000 and 100,000 kicks. Stars correspond to $\textrm{SU}(d)$
averaging while triangles correspond to $\textrm{SU}(2)$ averaging. The
Lyapunov exponent obtained from the classical dynamics of a single top is shown
by the dashed line. For the sake of later discussion we also included the mean
entanglements of eigenvectors of the evolution operator -- denoted by
rectangles.

In the case of $\textrm{SU}(2)$ averaging, for very low values of $k$, the
asymptotic values are high. With the increase of $k$ they decrease, reaching
minimum for $k \approx 1.5$, subsequently, with the onset of chaos, they
increase again and eventually saturate. The saturation value is a little below
the statistical limit (SL).

The $\textrm{SU}(d)$ averaging reveals almost no dependence of the asymptotic
value of entanglement on $k$. Nevertheless, there is also a tiny dip around $k
\approx 2$ and the entanglement for very low $k$ is a little bit higher than
that for the strongly chaotic regime.

The behavior of the mean eigenvector entanglement, also reveals a minimum
around $k \approx 2$. Remarkably the entanglement of eigenvectors for very high
$k$ is significantly smaller than asymptotic value of entanglement in this
regime.

\subsection{Initial behavior}

\begin{figure}[t]
\begin{center}
\includegraphics[width=0.5\textwidth]{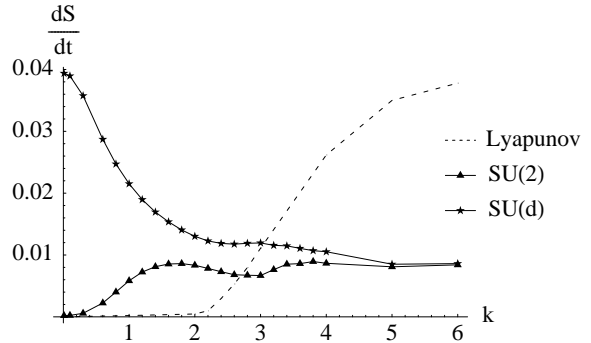}
\caption{Dependence of initial entanglement production rate on the chaoticity
parameter $k$. Results obtained after averaging over initial random product
states [the $\textrm{SU}(d)$ averaging] are denoted by stars and results due to
averaging over initial spin-coherent product states [the $\textrm{SU}(2)$
averaging] are denoted by triangles. The dashed line represents the rescaled
Lyapunov exponent of the classical dynamics of a single top. In the case of
random product states the highest initial entanglement growth corresponds to
very regular dynamics and diminishes with the increase of the chaoticity
parameter $k$ while in the case of spin-coherent product states it is very
regular dynamics that has the slowest initial entanglement growth.}
\label{comparison:start}
\end{center}
\end{figure}

In order to grasp quantitatively the initial behavior of entanglement, we fit a
line to the points representing short-time entanglement produced for certain
value of $k$. Although the character of initial entanglement growth is linear only
in the chaotic regime, while in the regular regime it is quadratic (see Sec. \ref{subsec:theory_initial}),
we perform the fitting in all regimes. This gives us sensible estimate of initial entanglement growth.
The fitting is done for points corresponding to first $15$ kicks.
Regression coefficients obtained in this way are shown in
Fig.~\ref{comparison:start} for both $\textrm{SU}(2)$ and $\textrm{SU}(d)$
averaging, together with the rescaled Lyapunov exponent of a single top. For
high values of $k$ both averaging methods give the same results, which is
caused by strong chaos which even during the first $15$ kicks is able to turn
spin-coherent states into completely random states. For low values of $k$ the
entanglement production rate obtained with  the $\textrm{SU}(d)$ averaging
outperforms that of the $\textrm{SU}(2)$ averaging. Furthermore it is
approximately four times higher than in the strong chaos regime. We shall now
move on to explaining these observations.

\section{Theoretical analysis of numerical results}

Before moving on to separate discussions on the asymptotic and initial behavior
of the entanglement production, let us first make a general remark which
applies to both cases. Looking at Fig.~\ref{comparison:asymptotic} and
Fig.~\ref{comparison:start} one can notice that when the chaoticity parameter
is large ($k > 5$) differences between the $\textrm{SU}(2)$ averaging and the
$\textrm{SU}(d)$ averaging disappear, both in results for the asymptotic and
the initial behavior. It means that there is no difference, whether we choose
as our initial states the ensemble of spin-coherent product states or the
ensemble of random product states. Disappearance of the difference between two
averaging methods for high $k$ is due to strong chaos which very quickly turns
initial spin-coherent states into random pure states. The distinction between
two averaging methods, however, is crucial in mixed and regular regimes.

\subsection{Entanglement of eigenvectors vs\ asymptotic values of entanglement
of evolved states}
\label{sec:eigenasymp}

Abstracting for a while from the case of the kicked tops, let $U$ be a unitary
operation acting in a Hilbert space $\mathcal{H}_1 \otimes \mathcal{H}_2$,
where dimensions of $\mathcal{H}_1$, $\mathcal{H}_2$ are respectively $d_1$ and
$d_2$. We denote by $|e_i \rangle$, $\phi_i$ for $i = 1 \dots d_1 d_2$ the
eigenvectors and eigenphases of the operator $U$. We assume here that the
spectrum is nondegenerate. The reduced density matrices of the eigenvectors
after tracing out the second subsystem are $\rho_i=\text{Tr}_2(|e_i\rangle
\langle e_i|)$. As information about entanglement of an eigenvector we use
again the linear entropy of its reduced density matrix: $S_i
=1-\text{Tr}(\rho_i^2)$. The mean entanglement of eigenvectors is given by
\begin{equation}
\label{eigen:entanglement}
\bar{S}_{\textrm{eigen}} =1-\frac{1}{d_1d_2}\sum\limits_i \text{Tr}(\rho_i^2).
\end{equation}

Assume that the initial state of the system is $|\psi\rangle$. After $n$
iterations of the operation $U$ the resulting state reads
\begin{equation}
|\psi(n) \rangle = \sum\limits_i \exp(i n \phi_i) \langle e_i | \psi \rangle |e_i\rangle.
\end{equation}
The reduced density matrix of the first subsystem corresponding to this state is
given by
\begin{equation}
\rho(n)= \sum\limits_{ij} \exp\left[in(\phi_i - \phi_j)\right] \langle e_i | \psi\rangle\langle \psi|e_j\rangle
 \text{Tr}_2(|e_i\rangle \langle e_j|),
\end{equation}
the linear entropy of which reads
\begin{equation}
\begin{array}{l}
S_{\rho(n)}= 1-\text{Tr}\left[\rho(n)^2\right]=
1-\text{Tr}\Bigl(\sum\limits_{ijkl} \exp[in(\phi_i - \phi_j +\phi_k - \phi_l)]\\
\langle e_i | \psi\rangle\langle \psi|e_j\rangle \langle e_k | \psi\rangle\langle \psi|e_l\rangle
\text{Tr}_2(|e_i\rangle \langle e_j|)\text{Tr}_2(|e_k\rangle \langle e_l|)\Bigr).
\end{array}
\end{equation}
For the asymptotic behavior of the linear entropy the important terms in the
above formula are those which survive after the time-averaging. These are the
terms in which phase factors disappear. Other terms oscillate and thus vanish
when averaged over time. The nonvanishing terms correspond to either $i=j$,
$k=l$ or $i=l$, $k=j$. Asymptotic value of linear entropy reads thus
\begin{equation}
\label{asymptotic}
\begin{array}{l}
S_{\textrm{asymp}} = 1-\text{Tr}\left(\sum\limits_{ij} |\langle e_i | \psi\rangle|^2 |\langle e_j | \psi \rangle|^2 \rho_i \rho_j\right) -
\\
\\
\text{Tr}\left(\sum\limits_{i\neq j}|\langle e_i | \psi\rangle|^2 |\langle e_j | \psi \rangle|^2\text{Tr}_2(|e_i\rangle\langle e_j |)
\text{Tr}_2(|e_j \rangle \langle e_i|)\right).
\end{array}
\end{equation}
We would like to stress here that the above formula is the result of the
averaging over time of the linear entropy itself. One could take a different
approach and first average over time the density matrix and than calculate the
linear entropy of the resulting density matrix. Following the second approach
one would not get the third term in the above formula. This difference is
important as it is certainly not the same to calculate the time-averaged
entanglement of the evolved state or to calculate the entanglement of the
time-averaged state. We argue that the first approach reveals better the
entangling properties of the system evolution. One must admit, however, that in
many cases the second approach will give qualitatively similar results.

We would like now to relate the above formula for the asymptotic entanglement
of an evolved state with the entanglement of eigenvectors. We start by proving
the following inequalities:
\begin{subequations}
\label{inequalities}
\begin{eqnarray}
\label{inequality1}
2\text{Tr}\left(\text{Tr}_2(|e_i\rangle\langle e_j|)\text{Tr}_2(|e_j\rangle\langle e_i|)\right) \leq \text{Tr}(\rho_i^2)+\text{Tr}(\rho_j^2)
\\
\label{inequality2}
2\text{Tr}(\rho_i \rho_j) \leq \text{Tr}(\rho_i^2)+\text{Tr}(\rho_j^2)
\end{eqnarray}
\end{subequations}
\emph{Proof.} We shall prove the first inequality. Let us decompose the
eigenvector $|e_i\rangle$ in a product basis $|e_i\rangle = \sum\limits_{n_1
n_2}C^i_{n_1n_2} |n_1 \rangle \otimes |n_2 \rangle$. We can write:
$\text{Tr}_2(|e_i\rangle\langle e_j|)=C^i C^{j \dagger}$. It follows that
\begin{eqnarray*}
\label{proof}
2\text{Tr}\left[\text{Tr}_2(|e_i\rangle\langle e_j|)\text{Tr}_2(|e_j\rangle\langle e_i|)\right] =
\\
= 2\text{Tr}(C^i C^{j \dagger} C^j C^{i \dagger})=2\text{Tr}(C^{i \dagger} C^i C^{j \dagger} C^j) \leq
\\
\leq \text{Tr}(C^{i \dagger}C^i C^{i \dagger}C^i) + \text{Tr}(C^{j \dagger} C^j C^{j \dagger} C^j)
= \text{Tr}(\rho_i^2)+ \text{Tr}(\rho_j^2).
\end{eqnarray*}
In the derivation above we have made use of the Cauchy-Schwarz inequality.
Assume that $A$ and $B$ are hermitian operators. The Cauchy-Schwarz inequality
than reads  $\text{Tr}(A^2)\text{Tr}(B^2) \geq \left[\text{Tr}(AB)\right]^2$. Taking into
account that $\left[\text{Tr}(A^2) + \text{Tr}(B^2)\right]^2 \geq
4\text{Tr}(A^2)\text{Tr}(B^2)$, one arrives at: $\text{Tr}(A^2) +
\text{Tr}(B^2) \geq 2\text{Tr}(AB)$. This explains the derivation above and
gives (\ref{inequality1}). The proof of the inequality (\ref{inequality2}) is
analogous.

We now return to the formula (\ref{asymptotic}). Using the inequalities
(\ref{inequalities}) we can write:
\begin{eqnarray*}
S_{\textrm{asymp}} = 1-\text{Tr}\left(\sum\limits_{ij} |\langle e_i | \psi\rangle|^2 |\langle e_j | \psi \rangle|^2 \rho_i \rho_j\right) -
\\
\\
\text{Tr}\left(\sum\limits_{i\neq j}|\langle e_i | \psi\rangle|^2 |\langle e_j | \psi \rangle|^2\text{Tr}_2(|e_i\rangle\langle e_j |)
\text{Tr}_2(|e_j \rangle \langle e_i|)\right) \geq \\
\geq  1-\text{Tr}\left(\sum\limits_{ij} |\langle e_i | \psi\rangle|^2 |\langle e_j | \psi \rangle|^2 \rho_i \rho_j\right) -
\\
\\
\text{Tr}\left(\sum\limits_{ij}|\langle e_i | \psi\rangle|^2 |\langle e_j | \psi \rangle|^2\text{Tr}_2(|e_i\rangle\langle e_j |)
\text{Tr}_2(|e_j \rangle \langle e_i|)\right) \geq \\
\geq 1 - 2\sum\limits_{i}|\langle e_i | \psi\rangle|^2
 \text{Tr}(\rho_i^2) =2  \sum\limits_{i}|\langle e_i | \psi\rangle|^2 S_i - 1,
\end{eqnarray*}
where $S_i$ is the linear entropy of the eigenstate $|e_i\rangle$.

We are interested in averaging over initial states $|\psi\rangle$ which are
either spin-coherent product states or random product states. Both averaging
procedures lead to
\begin{equation}
\label{eq:averaging}
\left\langle\left\langle|\langle e_i | \psi\rangle|^2\right\rangle\right\rangle_\psi = 1/d_1d_2.
\end{equation}
Finally we arrive at the following relation between mean
asymptotic value of entanglement of evolved states and mean entanglement of
eigenvectors:
\begin{equation}
\label{asymp:eigen}
\bar{S}_{\textrm{asymp}} \geq  2 \bar{S}_{\textrm{eigen}} -1.
\end{equation}
This inequality is valid for both
the $\textrm{SU}(2)$ averaging and the $\textrm{SU}(d)$ averaging. Actually, as the equation
(\ref{eq:averaging}) is valid for any ensemble of states which give a
resolution of identity, so is the equation (\ref{asymp:eigen}) valid in all
these cases and not only in the case of $\textrm{SU}(2)$ or $\textrm{SU}(d)$ averaging.
This inequality puts the lower bound on the mean asymptotic entanglement of
evolved states, thus a high entanglement of eigenvectors induce a high mean
asymptotic entanglement of evolved states.

One is not entitled, however, to claim that a low mean entanglement of
eigenvectors implies a low mean asymptotic entanglement of evolved states
(compare Fig.~\ref{comparison:asymptotic}). To stress once again the inadequate
information about entangling properties of a transformation obtained from
studying its eigenvector entanglement, recall that even when all eigenvectors
are product states, a transformation can have a non-zero entangling power (for
example the controlled-phase gate).

This is a good place to comment on our decision to take $j_1$ different from
$j_2$. If we took identical spins for the two tops the degeneracy of
eigenvectors would appear. The presence of degeneracy invalidate the inequality
(\ref{asymp:eigen}). In the presence of degeneracy estimating entangling power
from entanglement of eigenvectors can be even more misleading. As an example
consider a local transformation $U=U_1 \otimes U_2$, which obviously has the
zero entangling power. Let $|f_1\rangle \otimes |f_2 \rangle$ and $|g_1 \rangle
\otimes |g_2 \rangle$ be two degenerate product eigenstates. However,
eigenstates in this situation can equally well be taken
 as $(1/\sqrt{2}) \left(|f_1\rangle \otimes |f_2 \rangle +|g_1 \rangle \otimes
|g_2 \rangle \right)$ and $(1/\sqrt{2}) \left(|f_1\rangle \otimes |f_2 \rangle
-|g_1 \rangle \otimes |g_2 \rangle \right)$, which are entangled states.
Consequently calculating numerically eigenstates and their entanglement we
could arrive at the conclusion that the entanglement of eigenvectors is high
while the transformation itself has the zero entangling power. Nondegenerate
spectrum is thus an indispensable condition in studying relations between
entangling power of a transformation and entanglement of its eigenvectors.

\subsection{Asymptotic behavior}

When explaining the asymptotic value of entanglement of an evolved state in a
system of two strongly chaotic interacting subsystems one can formulate some
statistical predictions. One expects that states evolved in such systems tend
to generically random states. This means that writing an asymptotic state, in a
product basis:
\begin{equation}
|\psi_{\textrm{asymp}}\rangle = \sum\limits_{n_1 n_2} a_{n_1 n_2} |n_1\rangle \otimes |n_2 \rangle,
\end{equation}
one expects the coefficients $a_{n_1 n_2}$ to be independent random variables.
Depending on the symmetries of the system, one imposes different restrictions
on the coefficients arriving thus at different state ensembles. It is then
possible to calculate the linear entropy averaged over such ensembles.
\cite{zyczkowski2001,bandyopadhyay2002}

Choosing an ensemble of states which are uniformly distributed with respect to
the action of $\textrm{SU}(d_1 d_2)$ group, one arrives at averaged linear entropy \cite{zyczkowski2001}:
\begin{equation}
\label{RMT}
S_{SL} = 1-\frac{d_1+d_2}{d_1d_2+1}.
\end{equation}
Consequently, the above expression is equal to the average entanglement
(measured as linear entropy of reduced density matrix) of a generically random
state of the composite system. We shall use this expression, which we call
statistical limit (SL), as a reference for the asymptotic entanglement of
evolved states.

When dimensions of subspaces are similar $d_1 \approx d_2 \approx d$ and large
the statistical limit can be written in a simple form $S_{SL} \approx 1-2/d$.
Compare this with the limit imposed solely by the dimensionality of the
subspaces: $S^{\textrm{max}} = 1 - 1/d$, where $d$ is the dimension of the
smaller subspace.

\begin{figure}[t]
\begin{center}
\includegraphics[width=0.5\textwidth]{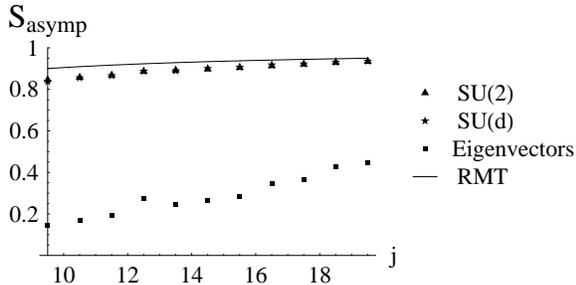}
\caption{Asymptotic entanglement of evolved states averaged over initial random
product states [the $\textrm{SU}(d)$ averaging, stars] and over initial
spin-coherent product states [the $\textrm{SU}(2)$ averaging, triangles) in
the strong chaos regime $k=6$, calculated for different spin magnitudes $j$.
Spins of the tops are chosen $j_1=j$ and $j_2=j+1/2$. Rectangles denote values of
mean entanglement of eigenvectors. The statistical limit (SL) for entanglement
is drawn with the solid line and corresponds to average entanglement of a
randomly chosen state of the composite system. A big discrepancy is visible
between the asymptotic entanglement of evolved states and the mean entanglement
of eigenvectors.} \label{chaos:asymptotic}
\end{center}
\end{figure}

\begin{figure}[t]
\begin{center}
\includegraphics[width=0.5\textwidth]{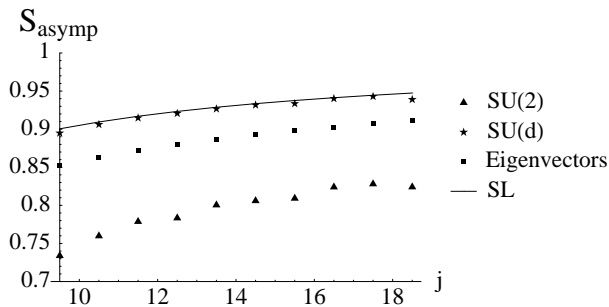}
\caption{Asymptotic entanglement of evolved states in the case of
averaging over initial random product states [$\textrm{SU}(d)$
averaging, stars] and in the case of  averaging over initial spin-coherent product states [$\textrm{SU}(2)$ averaging, triangles]
in the regular regime $k=0.01$, calculated for different spin magnitudes $j$.
Spins of the tops are chosen: $j_1=j$ and $j_2=j+1/2$.
Rectangles denote values of mean entanglement of eigenvectors.
Statistical limit (SL) for entanglement is drawn with a solid line and corresponds to
average entanglement of a randomly chosen state of the composite system. Thanks to inequality (\ref{asymp:eigen})
high entanglement of eigenvectors is an indicator of high asymptotic entanglement for both averaging procedures}
\label{regular:asymptotic}
\end{center}
\end{figure}

In Fig.~\ref{chaos:asymptotic} we compare this formula with the values of the
asymptotic entanglement for strong chaos $k=6$, for different spin magnitudes
of the two tops. There is no difference between values obtained for the
$\textrm{SU}(2)$ or the $\textrm{SU}(d)$ averaging, and both are a little below
the statistical limit (SL). Notice that the eigenvectors entanglement is hardly
of any use here in predictions of the asymptotic entanglement.

We noticed earlier that for low values of $k$ asymptotic values of entanglement
for the $\textrm{SU}(2)$ and the $\textrm{SU}(d)$ averaging are also very high.
In the light of the inequality (\ref{asymp:eigen}) this is due to the very high
entanglement of the eigenvectors of the evolution operator. Asymptotic
entanglements for $k=0.01$ are presented in Fig.~\ref{regular:asymptotic}
together with the entanglement of eigenvectors for different spin magnitudes.
The statistical limit (SL) (\ref{RMT}) is also plotted for comparison.

The agreement between the statistical limit (SL) and the asymptotic values
of entanglement of the $\textrm{SU}(d)$ averaging is purely accidental. This
underlines, however, that the regular regime can be as good (or better) in
generating high asymptotic values of entanglement for random states compared as
the strongly chaotic regime. This agrees with the results in \cite{gorin2001},
where different approach was taken -- no specific model of dynamics was
analyzed but instead the Hamiltonians of the subsystems where random matrices
chosen either from chaotic or regular ensembles.

It remains to be explained what is the reason for a very high entanglement of
eigenvectors for low values of $k$. This effect of high entanglement of
eigenvectors in the regular motion regime was also observed in the spin-kicked
rotor system \cite{tanaka1996} and a mechanism is similar here. We shall give a
qualitative explanation of this effect. For decoupled kicked tops $\epsilon=0$,
eigenvectors are product states. For low values of $k$ there are many states
that are nearly degenerate. Additionally, these product states are very close
to the product states of eigenvectors of the $J_y$ operator. Indeed for $k=0$
one time step of the evolution of the tops is described by the unitary
operator:
\begin{equation}
U_0=\exp(-ipJ_{y_1}-ipJ_{y_2}),
\end{equation}
so eigenvectors can be chosen as $|y_1,y_2 \rangle$, the product states of the
eigenvectors of $J_y$ operator. There are many degenerate eigenvectors of
$U_0$, for example $|-j,j\rangle, |-j+1,j-1\rangle, ... ,|j,-j\rangle$ -- these
are $2j+1$ states with the eigenphase $0$ (for simplicity we have assumed
$j_1=j_2$ and the value of the spin to be integer). Additionally there will be
also groups consisting of $2j,2j-1, \dots ,1$ degenerate states. The coupling
term $\exp(-\epsilon J_{z1} J_{z2})$, expanded to the first order in
$\epsilon$, will couple the state $|y_1,y_2\rangle$ with the states of the form
$|y_1-1,y_2+1\rangle, |y_1+1,y_2-1\rangle, |y_1+1,y_2+1\rangle,
|y_1-1,y_2-1\rangle$ (as $J_z$ is a sum of lowering and rising operators in the
basis of $J_y$ eigenvectors). A weak coupling between non-degenerate states
will not cause much change in the form of eigenvectors. Let $|v_i\rangle$ be
the set of degenerate eigenvectors of $U_0$. Due to the coupling, new
eigenvectors will be obtained by diagonalizing  a matrix of the approximately
following form:
\begin{equation*}
\left[\begin{array}{cccccc}
p        & \epsilon & 0        & \cdots \\
\epsilon & p        & \epsilon & \cdots \\
0        & \epsilon & p        & \cdots \\
\vdots   & \vdots   & \vdots   & \ddots
\end{array}\right].
\end{equation*}
Eigenvectors of such a matrix have large contributions from many different
vectors $v_i$, in our case it means that in every subspace of degenerate
product eigenvectors new highly entangled eigenvectors will emerge due to the
weak coupling.

\subsection{Initial behavior}
\label{subsec:theory_initial}
The initial entanglement growth in the system of coupled kicked tops can be
well understood with the help of the perturbation theory developed in
\cite{fujisaki2003,znidaric2003}.

It should be clarified here that, although the two papers
\cite{fujisaki2003,znidaric2003} contains similar results, the motivation and
the scope of their work is a bit different. The main motivation of the work by
Fujisaki et al. \cite{fujisaki2003} is to study entanglement production in
weakly coupled chaotic systems and as a model they consider the coupled kicked
tops. This is also the approach we take in this article. On the other hand,
\v{Z}nidari\v{c} and Prosen consider a problem of stability of quantum dynamics
of a composite system with respect to weak interaction between two subsystems.
They consider different quantities reflecting the stability of quantum motion
such as fidelity, reduced fidelity and purity fidelity, and analyze their
behavior under regular or chaotic dynamics for different times of evolution
(comprehensive study of the problem of stability of quantum dynamics can be found in \cite{znidaric2004}),
see also \cite{prosen2002, prosen2003, prosen2003b}).
Because the linear entropy ($S$), which is used as a measure of entanglement in
\cite{fujisaki2003} and in the present work, is related to the purity fidelity
($F_P$) by the formula $S =  1 - F_P$, there is a close relation between
results obtained in \cite{fujisaki2003} and \cite{znidaric2003}.

The essential quantity for understanding initial entanglement growth is the
product of the time correlation functions of uncoupled subsystems:
\begin{equation}
D(n,m)=C_1(n,m)C_2(n,m),
\end{equation}
where the correlation function of an uncoupled subsystem is defined
\begin{equation}
C_i(n,m)=\langle \hat{z}_i(n) \hat{z}_i(m)\rangle - \langle \hat{z}_i(n) \rangle \langle \hat{z}_i(m),
\end{equation}
and $\hat{z}_i(m) = J_{z_i}(m)/j_i$. The Heisenberg picture is used here and
$J_{z_i}(n)$ denotes the operator of projection of the angular momentum on the
$z$ axis at time $n$ for the subsystem $i$ (after the $n$-th kick of uncoupled
evolution).

The perturbation formula for the initial behavior of the linear entropy reads
\begin{equation}
S^{\textrm{perturb}}(t)= 2 \epsilon^2 j_1 j_2 \sum\limits_{n=1}^{t}\sum\limits_{m=1}^{t} D(n,m).
\label{perturbative:general}
\end{equation}
For strong chaos one can neglect the terms with $n \neq m$, as any correlation
is quickly washed out. For $n=m$ due to the chaotic character of the motion
$D(n,n) \approx 1/9$ and does not depend on $n$ \cite{fujisaki2003} (more
detailed analysis of the behavior of the correlation function can be found in
\cite{tanaka2002,fujisaki2003,prosen2002,znidaric2003}). Finally for strong
chaos one obtains that the initial growth of linear entropy  is linear, and the rate of the growth is given by:
\begin{equation}
\label{perturbative}
\frac{dS^{\textrm{perturb}}}{dt}= \frac{2}{9} \epsilon^2 j_1 j_2.
\end{equation}
This formula is valid for any initial state, actually not from the very
beginning of the evolution, but only after a short relaxation time when the
linear growth of the linear entropy emerges. In our case the linear growth
appears very quickly, after first few kicks.

\begin{figure}[t]
\begin{center}
\includegraphics[width=0.5\textwidth]{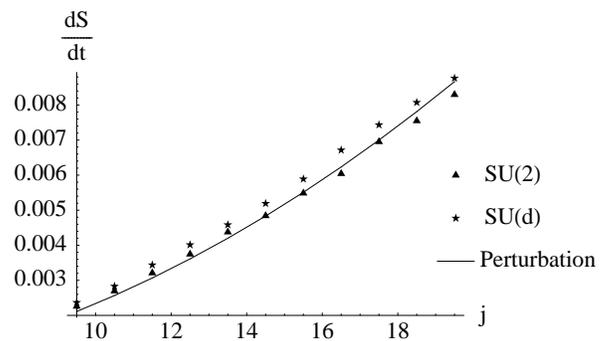}
\caption{Initial entanglement growth rates averaged over initial random product
states [the $\textrm{SU}(d)$ averaging -- stars] and initial entanglement growth
rates averaged over initial spin-coherent product states [the $\textrm{SU}(2)$
averaging -- triangles]. The results are obtained in strongly chaotic regime
$k=6$ and for different spin magnitudes $j$. The spins of the tops differ by
$1/2$: $j_1=j$ and $j2=j+1/2$. The prediction of perturbative formula
(\ref{perturbative}) is drawn with the solid line. Due to strong chaos there is
no significant difference between the spin-coherent and the random states.}
\label{chaos:start}
\end{center}
\end{figure}

The comparison of the above formula with the initial entanglement growth rate
corresponding to the $\textrm{SU}(2)$ and the $\textrm{SU}(d)$ averaging, in
the case of strong chaos $k=6$ is shown in Fig.~\ref{chaos:start}.
The agreement is good and, due to strong chaos, there is no significant
difference between two ensembles.

In the regular regime,
the formula ~(\ref{perturbative:general}), yields
initial quadratic increase of linear entropy, due to nonvanishing correlations
$C(n,m)$ for $n \neq m$ \cite{prosen2003,fujisaki2003}. In this regime there is
qualitatively different behavior of linear entropy increase between $\textrm{SU}(2)$ and $\textrm{SU}(d)$ averaging (see Fig.~\ref{comparison:start}).


In order to explain the difference we have to estimate values of the
correlation functions $C(n,m)$ averaged over the  $\textrm{SU}(2)$ and
$\textrm{SU}(d)$ ensembles of initial states. Notice the following formulas for
a particle with spin $j$:
\begin{subequations}
\label{su}
\begin{eqnarray}
\int d\mu_{\textrm{SU}(2)} \langle \psi | J_z^2| \psi \rangle &=& \frac{j(j+1)}{3}, \label{su2b} \\
\int d\mu_{\textrm{SU}(2)} \langle \psi | J_z| \psi \rangle^2 &=& \frac{j^2}{3},\label{su2s} \\
\int d\mu_{\textrm{SU}(d)} \langle \psi | J_z^2| \psi \rangle &=& \frac{j(j+1)}{3}, \label{sudb}\\
\int d\mu_{\textrm{SU}(d)} \langle \psi | J_z| \psi \rangle^2 &=& \frac{j}{6} \label{suds},
\end{eqnarray}
\end{subequations}
where $d=2j+1$ is the dimension of the particle's Hilbert space. Proofs of
these formulas are given in the appendix. The same equations can be written
when $J_z$ is replaced by either $J_y$ or $J_x$. If, however,  $\langle \psi
|J_z^2| \psi \rangle$ is replaced by a mixed term -- for example  $\langle \psi
|J_z J_x| \psi \rangle$ - the integrals (\ref{su2b}), (\ref{sudb}) vanish. The
same happens if in the integrals (\ref{su2s}), (\ref{su2s}) the term $\langle
\psi | J_z| \psi \rangle^2 $ is replaced by a mixed term - for example $\langle
\psi | J_z| \psi \rangle \langle \psi | J_x| \psi \rangle$.

In the most regular case ($k=0.01$) the evolution of a single top is mostly determined by its free
rotation around $y$ axis. Each period corresponds to a $\pi/2$ rotation.
Consequently in the Heisenberg picture operator $J_z$ evolves approximately in
the following way: $J_z(1) \approx J_x$,  $J_z(2) \approx -J_z$,  $J_z(3)
\approx -J_x$,
 $J_z(4) \approx J_z \ldots$.
The value of the correlation at equal times $C(n,n)$ is just the dispersion of $J_z(n)$ operator.
In the case of the $\textrm{SU}(2)$ averaging this will be proportional to the difference of integrals (\ref{su2b}) and (\ref{su2s})
and thus proportional to $1/j$. In contrast, $C(n,n)$ for the $\textrm{SU}(d)$
averaging is proportional to the difference of integrals (\ref{sudb}) and
(\ref{suds}) and thus proportional to $1$ -- the same order of magnitude as in
the chaotic case. The same fact will hold for correlation functions calculated for times differing
by even number of kicks, whereas correlation functions calculated for times differing by odd number of kicks will be zero.
As $D(n,m)$ is a product of two correlation functions for the
two tops, it will be approximately $j_1j_2$ times greater for the
$\textrm{SU}(d)$ averaging than for the $\textrm{SU}(2)$ averaging. In the case
we consider, this amounts to the ratio of initial entanglement production rate
of approximately $400$. This explains the strikingly different behavior of the
initial entanglement growth for two considered kinds of averaging in for the $k=0.01$ case.
Different character of entanglement production rate for coherent and random states was
also noticed in \cite{prosen2003}.

$\textrm{SU}(d)$ averaging reveals decrease in the initial entanglement growth rate with the increase of chaoticity parameter $k$
As the states we average over the ensemble of random states invariant under the action of $\textrm{SU}(d)$, the increase of parameter $k$ does not change the value of the dispersion of $J_z$ operator; thus $C(n,n)$ does not depend on $k$.
At the same time the increase of $k$ decreases correlation
for different times $C(n,m)$ ($n \neq m$), and consequently the increase of the chaoticity parameter
checks initial entanglement growth.

In the case of $\textrm{SU}(2)$ averaging, consequences of the increase of the chaoticity parameter are twofold.
On the one hand the increase of chaos decreases correlations $C(n,m)$ for $n \neq m$, and in this way
transforms initial entanglement growth character from quadratic to linear. On the other hand it drives
well-localized spin-coherent states into more smeared states with larger dispersion of $J_z$ operator, which
results in the increase of $C(n,n)$ function calculated at equal times. Depending on the specific values of parameters
of the system: $j_1$, $j_2$, $\epsilon$, one of theses factors may bare grater significance,
and it may happen, that the increase of chaos corresponds to
either the increase or decrease of initial entanglement growth (see for example numerical results in \cite{bandyopadhyay2003}
for different values of $\epsilon$ parameter).
We chose the parameters that reveal the increase of entanglement growth with growing $k$ for $\textrm{SU}(2)$ averaging
in order to contrast this with the behavior of entanglement growth in the case of $\textrm{SU}(d)$ averaging,
as for the latter case the increase of chaos \emph{always} diminishes initial entanglement growth.

\section{Conclusions}

In this paper, the problem of the interplay between entanglement production in
a quantum system and its chaotic properties was analyzed using the model of the
coupled kicked tops.

Entangling properties of the coupled kicked tops system were investigated by
observing the evolution of two different ensembles of product states.
Considerations of ensembles consisting either of product spin-coherent states
[$\textrm{SU}(2)$ averaging] or random product states [$\textrm{SU}(d]$
averaging) lead to qualitatively different results in terms of the initial
entanglement production rate and the asymptotic entanglement of evolved states.
The asymptotic values of entanglement are high in a strongly chaotic regime
(high kick strength), still they are even higher for a very regular regime (low
kick strength), and reach minimum for the parameters of the evolution that
correspond to the onset of chaos.

The $\textrm{SU}(2)$ averaging reveals strong dependence of the asymptotic
value of entanglement on the kick strength, while asymptotic values obtained
after the $\textrm{SU}(d)$ averaging are quite insensitive to variation of this
parameter. This is due to more classical nature of spin-coherent states, which
''feel'' more directly the transition form regular to chaotic dynamics in the
classical motion. In the case of the $\textrm{SU}(d)$ averaging, the classical
transition from regular to chaotic motion has only minor influence on the
asymptotic value of entanglement.

In order to gain a deeper understanding of the problems, the asymptotic
entanglement behavior was related to the mean entanglement of eigenvectors of
the transformation. We have proved an inequality relating these two quantities.
We have also pointed out cases where this quantities differ strongly and one is
not entitled to infer anything about one of them from the value of the other.

Finally we have studied the averaged initial entanglement production rate. The
striking difference between the $\textrm{SU}(2)$ and the $\textrm{SU}(d)$
averaging was explained with the help of the perturbation theory
\cite{znidaric2003,fujisaki2003}. In the regular regime spin-coherent product
states become entangled much slower than random product states, due to a significantly smaller value of
averaged variances of angular momentum components.

Regular to chaotic transition can indeed be observed in the global entangling
properties of the kicked tops. These manifestations, however, are different
depending on the ensembles of states considered. When considering random product states
the increase of the chaoticity parameter always diminishes initial entanglement growth.
For spin-coherent states increase of chaos results in two competing tendencies:
chaos drives coherent states into more delocalized states and thus helps entanglement growth,
but on the other hand it destroys time correlations in subsystems, which checks entanglement growth.
We chose parameters of the evolution in which first tendency prevailed and thus contrasted
the behavior of spin-coherent vs. random states.

When discussing asymptotic entanglement we have observed that
both purely regular and strongly chaotic regimes enjoy a very high
asymptotic entanglement. Reasons for high asymptotic entanglement in regular
and chaotic cases are different. In the chaotic case, chaotic dynamics in
subsystems allows for a coupling of arbitrary states and consequently an
initial product state can become highly entangled as the dynamics is able to
penetrate Hilbert space of the composite system with almost no constraints. In
the regular case evolution of a state is much more restricted, but in the
system of coupled kicked tops, due to a very high entanglement of eigenvectors
(see inequality \ref{asymp:eigen}) the asymptotic entanglement is also very
high.

\begin{acknowledgments}
We would like to thank T. Prosen and M. \v{Z}nidari\v{c} for informing us about
references \cite{prosen2002,znidaric2003,prosen2003}. This work was supported
by the EC grant QUPRODIS, contract No IST-2001-38877 and the Polish Ministry of
Scientific Research and Information Technology under the (solicited) grant No
PBZ-Min-008/P03/03.
\end{acknowledgments}

\appendix*
\section{Derivation of formulas (\ref{su})}
Formulas (\ref{su2b}, \ref{sudb}) can be explained using symmetry arguments. If
instead of $J_z^2$, we took $J^2 = J_x^2+ J_y^2 +J_z^2$, the mean value
$\langle \psi| J^2| \psi \rangle = j(j+1)$, for any state. Both
$\textrm{SU}(2)$ and $\textrm{SU}(d)$ averaging do not distinguish any
direction in space. Consequently averages of $J_x^2$, $J_y^2$ and $J_z^2$
should be equal. In order to sum up to $j(j+1)$ for $J^2$ operator each of
these averages must be equal $j(j+1)/3$

Let us now prove formula (\ref{su2s}). A spin-coherent state can be obtained as
a rotation of $|j\rangle$ state: $|\theta, \phi \rangle = R(\theta, \phi) |j
\rangle$. This allows us to write
\begin{eqnarray*}
\int d\mu_{\textrm{SU}(2)} \langle \psi |J_z| \psi \rangle^2 =
\frac{1}{4\pi}\int d \theta d\phi \sin(\theta)\langle \theta, \phi | J_z| \theta,\phi \rangle^2 = \\
=\frac{1}{4\pi}\int d \theta d\phi \sin(\theta)\langle -j| R(\theta, \phi)^\dagger  J_z R(\theta,\phi)|-j \rangle^2 =  \\
=\frac{1}{4\pi}\int d \theta d\phi \sin(\theta)\langle -j| J_z \cos(\theta) + J_x \sin(\theta)\cos(\phi) \\
- J_y\sin(\theta) \sin(\phi)|-j \rangle^2 = \frac{1}{4\pi}\int d \theta d\phi \sin(\theta) \cos^2(\theta) j^2 = \frac{j^2}{3}.
\end{eqnarray*}

The proof of formula (\ref{suds}) requires a little more effort. An arbitrary
state $|\psi\rangle$ can be written as: $|\psi\rangle = \sum\limits_{m=-j}^j
a_m |m \rangle$, where $a_m = x_m + i y_m$ are complex coefficients. The
$\textrm{SU}(d)$ averaging over $|\psi\rangle$ results in a random distribution
of $a_m$ coefficients with the probability distribution:
\begin{eqnarray*}
P(a_{-j}, \dots ,a_j)=P(x_{-j},y_{-j},\dots, x_j,y_j) = \\
=\frac{(2j)!}{\pi^{2j+1}} \delta(1-\sum\limits_{m=-j}^j x_{m}^2-\sum\limits_{m=-j}^j y_{m}^2).
\end{eqnarray*}
Marginal distributions can be calculated from the above formula. In what
follows we shall need the two lowest marginal distributions of $P$ i.e.\
$P_1(x)$ and $P_2(x,y)$ \cite{kus1988}:
\begin{eqnarray*}
P_1(x)=\frac{1}{\sqrt{\pi}}\frac{\Gamma(2j+1)}{\Gamma(2j+1/2)}(1-x^2)^{2j - \frac{1}{2}} \\
P_2(x,y)=\frac{(2j)!}{\pi}(1-x^2-y^2)^{2j-1}.
\end{eqnarray*}
Note also the following equalities:
\begin{subequations}
\label{p1p2}
\begin{eqnarray}
\label{p1}
\int dx \  x^4 P_1(x) = \frac{3}{8(j+1)(2j+1)} \\
\label{p2}
\int dxdy\  x^2y^2 P_2(x,y) = \frac{1}{8(j+1)(2j+1)}
\end{eqnarray}

We are now prepared to prove formula (\ref{suds}):
\begin{eqnarray*}
\int d\mu_{\textrm{SU}(d)} \langle \psi | J_z| \psi \rangle^2 = \int da_{-j} \dots da_j P(a_{-j},\dots a_j)\\
\left(\sum\limits_{n=-j}^j a_n^* \langle n|J_z  \sum\limits_{m=-j}^j a_m |m \rangle \right)^2 = \\
\int da_{-j} \dots da_j P(a_{-j},\dots a_j) \left(\sum\limits_{m=-j}^j |a_m|^2 m  \right)^2 = \\
\int da_{-j} \dots da_j P(a_{-j},\dots a_j) \sum\limits_{m=-j}^j \sum\limits_{n=-j}^j |a_m|^2 |a_n|^2 mn,
\end{eqnarray*}
only the terms with either $n=m$ or $n=-m$ contribute, as all others cancel
out:
\begin{eqnarray*}
\int da_{-j} \dots da_j P(a_{-j},\dots a_j) \sum\limits_{m=-j}^j |a_m|^4 m^2 - \\
\sum\limits_{m=-j}^j |a_m|^2 |a_{-m}|^2 m^2 =\\
\int dx_{-j}dy_{-j} \dots dx_{j}dy_j P(x_{-j},y_{-j},\dots, x_j,y_j) \\
\sum\limits_{m=-j}^j (x_m^4 + y_m^4+2x_m^2 y_m^2) m^2 -  \\
\sum\limits_{m=-j}^j (x_m^2+y_m^2)(x_{-m}^2 + y_{-m}^2)m^2=\\
\left(2\int dx \ x^4 P_1(x) - 2\int dxdy \ x^2y^2 P_2(x,y)\right) \sum\limits_{m=-j}^j m^2 =\\
\frac{1}{2(j+1)(2j+1)} \frac{j(j+1)(2j+1)}{3} = \frac{j}{6}.
\end{eqnarray*}

\end{subequations}


\begin{thebibliography}{27}
\expandafter\ifx\csname natexlab\endcsname\relax\def\natexlab#1{#1}\fi
\expandafter\ifx\csname bibnamefont\endcsname\relax
  \def\bibnamefont#1{#1}\fi
\expandafter\ifx\csname bibfnamefont\endcsname\relax
  \def\bibfnamefont#1{#1}\fi
\expandafter\ifx\csname citenamefont\endcsname\relax
  \def\citenamefont#1{#1}\fi
\expandafter\ifx\csname url\endcsname\relax
  \def\url#1{\texttt{#1}}\fi
\expandafter\ifx\csname urlprefix\endcsname\relax\def\urlprefix{URL }\fi
\providecommand{\bibinfo}[2]{#2}
\providecommand{\eprint}[2][]{\url{#2}}

\bibitem[{\citenamefont{Haake}(1991)}]{haake}
\bibinfo{author}{\bibfnamefont{F.}~\bibnamefont{Haake}},
  \emph{\bibinfo{title}{Quantum signatures of chaos}}
  (\bibinfo{publisher}{Springer-Verlag}, \bibinfo{year}{1991}).

\bibitem[{\citenamefont{Reichl}(1992)}]{reichl}
\bibinfo{author}{\bibfnamefont{L.~E.} \bibnamefont{Reichl}},
  \emph{\bibinfo{title}{The transition to chaos}}
  (\bibinfo{publisher}{Springer-Verlag}, \bibinfo{year}{1992}).

\bibitem[{\citenamefont{Furuya et~al.}(1997)\citenamefont{Furuya, Nemes, and
  Pellegrino}}]{furuya1997}
\bibinfo{author}{\bibfnamefont{K.}~\bibnamefont{Furuya}},
  \bibinfo{author}{\bibfnamefont{M.~C.} \bibnamefont{Nemes}}, \bibnamefont{and}
  \bibinfo{author}{\bibfnamefont{G.~Q.} \bibnamefont{Pellegrino}},
  \bibinfo{journal}{Phys. Rev. Lett.} \textbf{\bibinfo{volume}{80}},
  \bibinfo{pages}{5524} (\bibinfo{year}{1997}).

\bibitem[{\citenamefont{Miller and Sarkar}(1998)}]{miller1998}
\bibinfo{author}{\bibfnamefont{P.~A.} \bibnamefont{Miller}} \bibnamefont{and}
  \bibinfo{author}{\bibfnamefont{S.}~\bibnamefont{Sarkar}},
  \bibinfo{journal}{Phys. Rev. E} \textbf{\bibinfo{volume}{60}},
  \bibinfo{pages}{1542} (\bibinfo{year}{1998}).

\bibitem[{\citenamefont{Angelo et~al.}(1999)\citenamefont{Angelo, Furuya,
  Nemes, and Pellegrino}}]{angelo1999}
\bibinfo{author}{\bibfnamefont{R.~M.} \bibnamefont{Angelo}},
  \bibinfo{author}{\bibfnamefont{K.}~\bibnamefont{Furuya}},
  \bibinfo{author}{\bibfnamefont{M.~C.} \bibnamefont{Nemes}}, \bibnamefont{and}
  \bibinfo{author}{\bibfnamefont{G.~Q.} \bibnamefont{Pellegrino}},
  \bibinfo{journal}{Phys. Rev. E} \textbf{\bibinfo{volume}{60}},
  \bibinfo{pages}{5407} (\bibinfo{year}{1999}).

\bibitem[{\citenamefont{A.Tanaka et~al.}(2002)\citenamefont{A.Tanaka, Fujisaki,
  and Miyadera}}]{tanaka2002}
\bibinfo{author}{\bibnamefont{A.Tanaka}},
  \bibinfo{author}{\bibfnamefont{H.}~\bibnamefont{Fujisaki}}, \bibnamefont{and}
  \bibinfo{author}{\bibfnamefont{T.}~\bibnamefont{Miyadera}},
  \bibinfo{journal}{Phys. Rev. E} \textbf{\bibinfo{volume}{66}},
  \bibinfo{pages}{045201(R)} (\bibinfo{year}{2002}).

\bibitem[{\citenamefont{Fujisaki et~al.}(2003)\citenamefont{Fujisaki, Miyadera,
  and Tanaka}}]{fujisaki2003}
\bibinfo{author}{\bibfnamefont{H.}~\bibnamefont{Fujisaki}},
  \bibinfo{author}{\bibfnamefont{T.}~\bibnamefont{Miyadera}}, \bibnamefont{and}
  \bibinfo{author}{\bibfnamefont{A.}~\bibnamefont{Tanaka}},
  \bibinfo{journal}{Phys. Rev. E} \textbf{\bibinfo{volume}{67}},
  \bibinfo{pages}{066201} (\bibinfo{year}{2003}).

\bibitem[{\citenamefont{Prosen and Seligman}(2002)}]{prosen2002}
\bibinfo{author}{\bibfnamefont{T.}~\bibnamefont{Prosen}} \bibnamefont{and}
  \bibinfo{author}{\bibfnamefont{T.~H.} \bibnamefont{Seligman}},
  \bibinfo{journal}{J. Phys. A: Math. Gen.} \textbf{\bibinfo{volume}{35}},
  \bibinfo{pages}{4707} (\bibinfo{year}{2002}).

\bibitem[{\citenamefont{Prosen et~al.}(2003{\natexlab{a}})\citenamefont{Prosen,
  Seligman, and \v{Z}nidari\v{c}}}]{prosen2003}
\bibinfo{author}{\bibfnamefont{T.}~\bibnamefont{Prosen}},
  \bibinfo{author}{\bibfnamefont{T.~H.} \bibnamefont{Seligman}},
  \bibnamefont{and}
  \bibinfo{author}{\bibfnamefont{M.}~\bibnamefont{\v{Z}nidari\v{c}}},
  \bibinfo{journal}{Phys. Rev. A} \textbf{\bibinfo{volume}{67}},
  \bibinfo{pages}{042112} (\bibinfo{year}{2003}{\natexlab{a}}).

\bibitem[{\citenamefont{\v{Z}nidari\v{c} and Prosen}(2003)}]{znidaric2003}
\bibinfo{author}{\bibfnamefont{M.}~\bibnamefont{\v{Z}nidari\v{c}}}
  \bibnamefont{and} \bibinfo{author}{\bibfnamefont{T.}~\bibnamefont{Prosen}},
  \bibinfo{journal}{J. Phys. A: Math. Gen.} \textbf{\bibinfo{volume}{36}},
  \bibinfo{pages}{2463} (\bibinfo{year}{2003}).

\bibitem[{\citenamefont{Wang et~al.}(2003)\citenamefont{Wang, Ghose, Sanders,
  and Hu}}]{wang2003}
\bibinfo{author}{\bibfnamefont{X.}~\bibnamefont{Wang}},
  \bibinfo{author}{\bibfnamefont{S.}~\bibnamefont{Ghose}},
  \bibinfo{author}{\bibfnamefont{B.}~\bibnamefont{Sanders}}, \bibnamefont{and}
  \bibinfo{author}{\bibfnamefont{B.}~\bibnamefont{Hu}},
  \bibinfo{journal}{quant-ph/0312047}  (\bibinfo{year}{2003}).

\bibitem[{\citenamefont{Bandyopadhyay and
  Lakshminarayan}(2004)}]{bandyopadhyay2003}
\bibinfo{author}{\bibfnamefont{J.~N.} \bibnamefont{Bandyopadhyay}}
  \bibnamefont{and}
  \bibinfo{author}{\bibfnamefont{A.}~\bibnamefont{Lakshminarayan}},
  \bibinfo{journal}{Phys. Rev. E} \textbf{\bibinfo{volume}{69}},
  \bibinfo{pages}{016201} (\bibinfo{year}{2004}).

\bibitem[{\citenamefont{Lakshminarayan}(2001)}]{lakshminarayan2001}
\bibinfo{author}{\bibfnamefont{A.}~\bibnamefont{Lakshminarayan}},
  \bibinfo{journal}{Phys. Rev. E} \textbf{\bibinfo{volume}{64}},
  \bibinfo{pages}{036207} (\bibinfo{year}{2001}).

\bibitem[{\citenamefont{Zanardi et~al.}(2000)\citenamefont{Zanardi, Zalka, and
  Faoro}}]{Zanardi2000}
\bibinfo{author}{\bibfnamefont{P.}~\bibnamefont{Zanardi}},
  \bibinfo{author}{\bibfnamefont{C.}~\bibnamefont{Zalka}}, \bibnamefont{and}
  \bibinfo{author}{\bibfnamefont{L.}~\bibnamefont{Faoro}},
  \bibinfo{journal}{Phys. Rev. A} \textbf{\bibinfo{volume}{62}},
  \bibinfo{pages}{030301(R)} (\bibinfo{year}{2000}).

\bibitem[{\citenamefont{Werner}(1989)}]{werner1989}
\bibinfo{author}{\bibfnamefont{R.~F.} \bibnamefont{Werner}},
  \bibinfo{journal}{Phys. Rev. A} \textbf{\bibinfo{volume}{40}},
  \bibinfo{pages}{4277} (\bibinfo{year}{1989}).

\bibitem[{\citenamefont{Bennett et~al.}(1996)\citenamefont{Bennett, Bernstein,
  Popescu, and Schumacher}}]{bennett1989}
\bibinfo{author}{\bibfnamefont{C.~H.} \bibnamefont{Bennett}},
  \bibinfo{author}{\bibfnamefont{H.~J.} \bibnamefont{Bernstein}},
  \bibinfo{author}{\bibfnamefont{S.}~\bibnamefont{Popescu}}, \bibnamefont{and}
  \bibinfo{author}{\bibfnamefont{B.}~\bibnamefont{Schumacher}},
  \bibinfo{journal}{Phys. Rev. A} \textbf{\bibinfo{volume}{53}},
  \bibinfo{pages}{2046} (\bibinfo{year}{1996}).

\bibitem[{\citenamefont{Vidal}(2000)}]{vidal2000}
\bibinfo{author}{\bibfnamefont{G.}~\bibnamefont{Vidal}}, \bibinfo{journal}{J.
  Mod. Opt} \textbf{\bibinfo{volume}{47}}, \bibinfo{pages}{355}
  (\bibinfo{year}{2000}).

\bibitem[{\citenamefont{Bandyopadhyay and
  Lakshminarayan}(2002)}]{bandyopadhyay2002}
\bibinfo{author}{\bibfnamefont{J.~N.} \bibnamefont{Bandyopadhyay}}
  \bibnamefont{and}
  \bibinfo{author}{\bibfnamefont{A.}~\bibnamefont{Lakshminarayan}},
  \bibinfo{journal}{Phys. Rev. Lett.} \textbf{\bibinfo{volume}{89}},
  \bibinfo{pages}{060402} (\bibinfo{year}{2002}).

\bibitem[{\citenamefont{Nielsen and Chuang}(2000)}]{nielsen}
\bibinfo{author}{\bibfnamefont{M.~A.} \bibnamefont{Nielsen}} \bibnamefont{and}
  \bibinfo{author}{\bibfnamefont{I.~L.} \bibnamefont{Chuang}},
  \emph{\bibinfo{title}{Quantum Computing and Quantum Information}}
  (\bibinfo{publisher}{Cambridge University Press}, \bibinfo{year}{2000}).

\bibitem[{\citenamefont{Haake et~al.}(1987)\citenamefont{Haake, Ku\'s, and
  Scharf}}]{hks87a}
\bibinfo{author}{\bibfnamefont{F.}~\bibnamefont{Haake}},
  \bibinfo{author}{\bibfnamefont{M.}~\bibnamefont{Ku\'s}}, \bibnamefont{and}
  \bibinfo{author}{\bibfnamefont{R.}~\bibnamefont{Scharf}},
  \bibinfo{journal}{Z. Phys. B} \textbf{\bibinfo{volume}{65}},
  \bibinfo{pages}{381} (\bibinfo{year}{1987}).

\bibitem[{\citenamefont{\.Zyczkowski and Sommers}(2001)}]{zyczkowski2001}
\bibinfo{author}{\bibfnamefont{K.}~\bibnamefont{\.Zyczkowski}}
  \bibnamefont{and} \bibinfo{author}{\bibfnamefont{H.~J.}
  \bibnamefont{Sommers}}, \bibinfo{journal}{J. Phys. A: Math. Gen.}
  \textbf{\bibinfo{volume}{34}}, \bibinfo{pages}{7111} (\bibinfo{year}{2001}).

\bibitem[{\citenamefont{Po\'zniak et~al.}(1998)\citenamefont{Po\'zniak,
  \.Zyczkowski, and Ku\'s}}]{pozniak2000}
\bibinfo{author}{\bibfnamefont{M.}~\bibnamefont{Po\'zniak}},
  \bibinfo{author}{\bibfnamefont{K.}~\bibnamefont{\.Zyczkowski}},
  \bibnamefont{and} \bibinfo{author}{\bibfnamefont{M.}~\bibnamefont{Ku\'s}},
  \bibinfo{journal}{J. Phys. A: Math. Gen.} \textbf{\bibinfo{volume}{31}},
  \bibinfo{pages}{1059} (\bibinfo{year}{1998}).

\bibitem[{\citenamefont{Gorin and Seligman}(2002)}]{gorin2001}
\bibinfo{author}{\bibfnamefont{T.}~\bibnamefont{Gorin}} \bibnamefont{and}
  \bibinfo{author}{\bibfnamefont{T.~H.} \bibnamefont{Seligman}},
  \bibinfo{journal}{J. Opt. B: Quantum Semiclass. Opt.}
  \textbf{\bibinfo{volume}{4}}, \bibinfo{pages}{S386} (\bibinfo{year}{2002}),
  \urlprefix\url{quant-ph/0112030}.

\bibitem[{\citenamefont{Tanaka}(1996)}]{tanaka1996}
\bibinfo{author}{\bibfnamefont{A.}~\bibnamefont{Tanaka}}, \bibinfo{journal}{J.
  Phys. A: Math. Gen.} \textbf{\bibinfo{volume}{29}}, \bibinfo{pages}{5475}
  (\bibinfo{year}{1996}).

\bibitem[{\citenamefont{\v{Z}nidari\v{c}}(2004)}]{znidaric2004}
\bibinfo{author}{\bibfnamefont{M.}~\bibnamefont{\v{Z}nidari\v{c}}},
  \bibinfo{journal}{quant-ph/0406123}  (\bibinfo{year}{2004}).

\bibitem[{\citenamefont{Prosen et~al.}(2003{\natexlab{b}})\citenamefont{Prosen,
  Seligman, and \v{Z}nidari\v{c}}}]{prosen2003b}
\bibinfo{author}{\bibfnamefont{T.}~\bibnamefont{Prosen}},
  \bibinfo{author}{\bibfnamefont{T.~H.} \bibnamefont{Seligman}},
  \bibnamefont{and}
  \bibinfo{author}{\bibfnamefont{M.}~\bibnamefont{\v{Z}nidari\v{c}}},
  \bibinfo{journal}{Prog. Theo. Phys. Supp.} \textbf{\bibinfo{volume}{150}},
  \bibinfo{pages}{200} (\bibinfo{year}{2003}{\natexlab{b}}),
  \urlprefix\url{quant-ph/0304104}.

\bibitem[{\citenamefont{Ku\'s et~al.}(1988)\citenamefont{Ku\'s, Mostowski, and
  Haake}}]{kus1988}
\bibinfo{author}{\bibfnamefont{M.}~\bibnamefont{Ku\'s}},
  \bibinfo{author}{\bibfnamefont{J.}~\bibnamefont{Mostowski}},
  \bibnamefont{and} \bibinfo{author}{\bibfnamefont{F.}~\bibnamefont{Haake}},
  \bibinfo{journal}{J. Phys. A: Math. Gen.} \textbf{\bibinfo{volume}{21}},
  \bibinfo{pages}{L1073} (\bibinfo{year}{1988}).

\end{thebibliography}
\end{document}